\documentclass[12pt]{article}
\usepackage{jheppub}
\usepackage{graphicx} 
\usepackage{amsfonts}
\usepackage{amsmath}
\usepackage{amssymb}
\usepackage{bbm,bm}
\usepackage{datetime}
\usepackage{footmisc}
\usepackage[dvipsnames]{xcolor}
\let\normalcolor\relax

\usepackage{slashed}
\usepackage{physics}
 \usepackage{float}

\definecolor{darkgreen}{rgb}{0,0.6,0}

\newcommand{\be}{\begin{equation}}
\newcommand{\ee}{\end{equation}}

\def\l{\lambda}

\def\Tr{\mathrm{Tr}}

\preprint{APCTP Pre2024 - 013}

\title{Chiral Separation Effect
from Holographic QCD}

\author[a]{Domingo Gallegos}
\affiliation[a]{Facultad de Ciencias, Universidad Nacional Aut\'onoma de M\'exico, Investigaci\'on Cient\'ifica C.U., 04510 Coyoacan, Ciudad de Mexico, Mexico }
\author[b,c]{Matti J\"arvinen}
\affiliation[b]{Asia Pacific Center for Theoretical Physics, Pohang, 37673, Korea}
\affiliation[c]{Department of Physics, Pohang University of Science and Technology, Pohang, 37673, Korea}
\author[d,e]{Eamonn Weitz}
\affiliation[d]{SUBATECH, IMT Atlantique, 4 Rue Alfred Kastler, La Chantrerie BP 20722, 44307 Nantes, France}
\affiliation[e]{Nantes Université, 2 Rue de la Houssinière, BP 92208,
44322 Nantes, France}
\emailAdd{matti.jarvinen@apctp.org}
\emailAdd{eamonn.weitz@subatech.in2p3.fr}

\abstract{
We analyze the chiral separation effect (CSE) in QCD by using the gauge/gravity duality. In QCD, this effect arises from a combination of chiral anomalies and the axial $U(1)$ anomaly. Due to the axial gluon anomaly, the value of the CSE conductivity is not determined by the anomalies of QCD but receives radiative corrections, which leads to nontrivial dependence on temperature and density. To analyze this dependence, we use different variants of the V-QCD, a complex holographic model, carefully fitted to QCD data. 
We find our results for the anomalous CSE conductivity at small chemical potential and nonzero temperature to be in good qualitative agreement with recent results from lattice QCD simulations. We furthermore give predictions for the behavior of the conductivity at finite (vectorial and axial) chemical potentials.  
}

\begin{document}

\maketitle

\section{Introduction}

Anomalies in field theory leave their imprint on the transport properties of the matter they describe. They can give rise exotic transport phenomena, which may lead to observable effects in real-world systems. Examples of such systems include the Dirac and Weyl semimetals, and perhaps the quark-gluon plasma, generated in heavy-ion collisions~\cite{Kharzeev:2024zzm}. In this latter context, the best known exotic phenomenon is the chiral magnetic effect (CME)~\cite{Kharzeev:2007jp,Fukushima:2008xe}, where the combination of a magnetic field and chirality imbalance (i.e., difference of number densities of left and right handed quarks) leads to a net current in the direction of the magnetic field. A related phenomenon is the chiral separation effect (CSE)~\cite{Metlitski:2005pr,Son:2009tf}, where a magnetic field at finite density leads to the generation of a chiral current. The chiral magnetic wave (CMW)~\cite{Kharzeev:2010gd,Burnier:2012ae} is then a collective hydrodynamic 
excitation that can arise from the interplay between vector and axial density fluctuations, coupled together by anomalous transport effects (also in the presence of an external magnetic field). Interestingly, charge asymmetries potentially related to these effects have indeed been observed in heavy-ion collisions~\cite{Belmont:2014lta,STAR:2015wza,STAR:2021mii}.  These effects may also be studied by using lattice QCD~\cite{Buividovich:2009wi,Yamamoto:2011gk,Yamamoto:2011ks,Buividovich:2013hza,Brandt:2023wgf,Brandt:2024wlw}.

The CME and the CSE are driven by chiral anomalies. That is, the coupling of the field theory in question to external fields leads to a non-conservation of the chiral currents. Associated conservation laws are then violated in a specific way which can be computed from triangle diagrams. This leads to the anomalous conductivities associated with these effects having specific ``universal'' values, which are determined by the anomalies~\cite{Fukushima:2008xe,Son:2009tf,Neiman:2010zi,Jensen:2012kj}. However, in interesting systems, the fermions are typically interacting via a strongly coupled gauge theory, which complicates the physical picture. Namely, the dynamical gauge fields also give rise to an anomaly. In the case of QCD, the theory describing quark-gluon plasma, this is the U(1) axial anomaly: the axial current conservation is violated in QCD per se. As it turns out, the radiative corrections from dynamical gauge fields affect some of the anomalous conductivities, driving them away from the universal values~\cite{Golkar:2012kb,Jensen:2013vta}. 

The gauge-gravity duality is a useful tool to study anomalous transport in strongly coupled systems. From the dual perspective, chiral symmetry is enhanced to a higher dimensional gauge symmetry, and the corresponding gauge fields describe the dynamics of the chiral currents. The chiral anomalies are then implemented through Chern-Simons terms of the gauge fields. In such setups, universal values for the anomalous conductivities have been demonstrated to hold quite generally~\cite{Gynther:2010ed,Amado:2011zx,Landsteiner:2011cp,Gursoy:2014boa,Gursoy:2014ela,Grozdanov:2016ala}. 
Be that as it may, the anomalies of the dynamical gauge fields arise as some of the higher dimensional gauge fields acquire a mass. In the case of QCD, the massive field is the U(1) axial gauge field. As usual, gauge symmetry does not allow one to add a hard mass term, but the mass can be generated by other means. One possibility, which is the mechanism appearing in gravity duals of QCD~\cite{Schwarz:2001sf,Sakai:2004cn,Casero:2007ae,Arean:2013tja,Arean:2016hcs} is the St\"uckelberg mechanism where the mass arises due to coupling of the gauge field to an additional scalar field~\cite{Jimenez-Alba:2014iia}.\footnote{In QCD, this field is dual to the $F\wedge F$ operator and the corresponding source is the $\theta$-angle.} The radiative corrections to anomalous conductivities in such setups have been studied in~\cite{Jimenez-Alba:2014iia,Jimenez-Alba:2015awa,Gallegos:2018ozs,Rai:2023nxe}. In addition, effects of additional mixed gauge-gravitational anomaly have been studied in~\cite{Landsteiner:2011iq,Megias:2013joa}. For systematics of hydrodynamics and transport in the presence of anomalies, see e.g.,~\cite{Son:2009tf,Neiman:2010zi,Kalaydzhyan:2011vx,Megias:2013joa,Bu:2018psl,Bu:2019mow,Ammon:2020rvg}. 

We note however that the majority of the analysis of anomalous conductivities within the gauge/gravity duality has been carried out in generic gravity backgrounds that are not adjusted to be dual to any specific field theory such as QCD. That is, while the chiral symmetries of some of the models are the same as in QCD, the models disagree with lattice or experimental data for QCD.
Even though some work on anomalous transport in specific holographic models for QCD exists~\cite{Yee:2009vw,Rebhan:2009vc,Gorsky:2010xu,Ballon-Bayona:2012qnu,Fukushima:2021got}, radiative corrections due to dynamical gluons  have not been addressed.

An effective approach to holographic QCD uses generic, ``bottom-up'' models possessing a large number of free parameters, which are determined through comparison to QCD data~\cite{Gursoy:2007cb,Gursoy:2007er,Gubser:2008ny,DeWolfe:2010he}. These models include improved holographic QCD~\cite{Gursoy:2007cb,Gursoy:2007er} and its extension to include a fully backreacted quark sector, the V-QCD model~\cite{Jarvinen:2011qe}. In particular, these models can be fitted~\cite{Gursoy:2009jd,Jokela:2018ers} to agree precisely with lattice data for QCD thermodynamics at finite temperature and low density. Therefore, they have the potential to produce realistic predictions for various observables in this region, which are relevant for heavy-ion collisions.

 Indeed, our goal here is to carry out an example of such an analysis, using V-QCD, focusing on anomalous transport in the deconfined quark-gluon plasma phase. 

The complete V-QCD setup~\cite{Bigazzi:2005md,Casero:2007ae,Jarvinen:2011qe} is consistent with global anomalies of QCD, including the U$(1)$ axial anomaly~\cite{Arean:2013tja,Arean:2016hcs}. For this model, the fit to lattice data for QCD thermodynamics has been done in~\cite{Jokela:2018ers,Ishii:2019gta}, and the model has been compared to QCD spectrum in~\cite{Amorim:2021gat,Jarvinen:2022gcc}. Thus, the remaining task is to carry out a fluctuation analysis in order to compute the anomalous conductivities, which we will do in this article.

Our computation assumes zero quark masses so that the high-temperature quark-gluon plasma phase is chirally symmetric. We then compare our results at low density to the recent lattice analysis~\cite{Brandt:2023wgf} and also give predictions at higher density. We turn on both vectorial (i.e., baryon number) and axial chemical potentials. Note that due to the axial anomaly there is no conserved charge corresponding to the axial chemical potential. Therefore this chemical potential does not have a clear thermodynamical interpretation, but should rather be seen as an additional coupling in the QCD action which runs as a function of the energy scale. Nevertheless, it is possibly that parity-violating bubbles form in heavy-ion collisions~\cite{Kharzeev:1998kz,STAR:2009wot}, so that an effective axial chemical potential is created locally.

The article is organized as follows. In Sec.~\ref{sec:vqcd}, we give a brief review of the V-QCD model. In Sec.~\ref{sec:conductivities}, we show how to compute anomalous conductivities via fluctuation analysis in V-QCD. In Sec.~\ref{sec:results}, we present our numerical results for the chiral separation effect and compare to results from the lattice. We conclude in Sec.~\ref{sec:conclusions}. The various Appendices contain additional details on the definition of the holographic model, its associated equations of motion and the extraction of observables.

\section{Holographic Model: V-QCD}\label{sec:vqcd}
The class of bottom-up models we choose to work with is V-QCD \cite{Jarvinen:2011qe,Alho:2012mh}, which is based on the combination of two frameworks: Improved Holographic QCD (IHQCD) \cite{Gursoy:2007cb,Gursoy:2007er}, describing the gluonic sector as well as a setup based on a Dirac-Born-Infeld (DBI) action, which models the quark sector \cite{Bigazzi:2005md,Casero:2007ae}. The two sectors are fully backreacted in the Veneziano limit where both the number of colors, $N_c$ and the number of flavors, $N_f$ are taken to be large:
\begin{align}
N_{c}\rightarrow\infty\ ,\quad N_{f}&\rightarrow\infty\ ,\quad\mathrm{with}\ \ \frac{N_{f}}{N_{c}}=x\ \ \mathrm{fixed}\ .
\end{align}  
The ``V'' in the name of the models refers to the use of this limit when the model is defined.  That is, in principle, one would like to derive the model from a concrete string theory background in this limit. However, since finding such a derivation is extremely challenging, string theory will be used merely as a guideline on how the Ansatz for the model action is chosen. 
Consequently, we switch to a bottom-up approach, 
where we set $N_c=N_f=3$ and the various sectors in the model are directly fitted to available QCD data (mostly lattice data for the thermodynamics of QCD) so that only the model action is inspired by string theory. We will now discuss the content of the model in more detail. See the recent review~\cite{Jarvinen:2021jbd} for a more detailed discussion.

\subsection{The action of the model}

The fields that we will be concerned with from the IHQCD sector are the following:
\begin{itemize}
    \item The dilaton, $\lambda=e^{\Phi}$, which is dual to the scalar gluonic operator, $\Tr\, G_{\mu\nu}G^{\mu\nu}$ and therefore sources the 't Hooft coupling in Yang-Mills theory.
    \item The metric $g_{MN}$, which is dual to the energy-momentum tensor, $T_{\mu\nu}$ of QCD. In this case, the source is the metric of the field theory, which we keep fixed as the Minkowski metric.  
\end{itemize}
Our notation is such that capital Latin indices run over all five dimensions of the bulk theory, whereas the Greek indices denote the Lorentz indices of the boundary theory.  We will also consider an axion field arising from this sector that we will discuss separately below.

The gluonic part of our action is then based on Einstein-dilaton theory,
 \begin{equation}
 S_{g}=M_{p}^{3}N_{c}^{2}\int d^{5}x\sqrt{-g}\Bigg(R-\frac{4}{3}\frac{(\partial_{M}\lambda)^{2}}{\lambda^{2}}+V_{g}(\lambda)\Bigg)\ ,
 \end{equation}
 where $M_{p}$ is the Planck mass. Notice that the non-trivial potential for the dilaton, $V_{g}(\lambda)$, is introduced to break the conformal symmetry and to mimic the nontrivial renormalization group flow of the coupling in QCD~\cite{Gursoy:2007cb,Gursoy:2007er}. We will discuss explicitly below our choice of this function, as well as various other functions appearing in the actions for the other sectors of the model.

Quarks are accounted for in the holographic setup by embedding $N_{f}$ space-filling $D4$-branes and $N_{f}$ space-filling $\bar{D}4$-branes in the 5D geometry~\cite{Bigazzi:2005md,Casero:2007ae}. In this article, we restrict ourselves to the (exactly) chirally symmetric solutions so that the quark mass is also taken to be vanishing\footnote{Though we do not include it here, chiral symmetry breaking or quark mass can be introduced in the V-QCD setup through a tachyon field~\cite{Bigazzi:2005md,Casero:2007ae,Jarvinen:2011qe}.}. Moreover, for our purposes, it is enough to consider the Abelian sector of the flavor fields, i.e., fields proportional to unit matrices in flavor space.  In this case, the low-lying fields on the branes are two gauge fields: $A_{L,\mu}$ on the flavor branes and $A_{R,\mu}$ on the anti-flavor branes, which are dual to the currents $\bar \psi \gamma_\mu(1\pm \gamma_5)\psi$ in QCD (with a sum over flavors included implicitly).
For our purposes, it will be convenient to rewrite the gauge fields in the basis of vector and axial vector fields:
 \begin{align}
 A_{L,M}&=V_{M}+A_{M}\ ,
\\A_{R,M}&=V_{M}-A_{M}\ .
 \end{align}
In this basis, the sources for (the temporal components of) the vectorial and axial gauge fields are the vectorial and axial quark chemical potentials, $\mu_V$ and $\mu_A$, respectively, as we will discuss in more detail below.
 
The flavor action is the effective  
DBI term
 \begin{equation} \label{eq:DBI}
 S_{f}=-\frac{xM_{p}^{3}N_{c}^{2}}{2}\int d^{5}xV_{f}(\lambda)\Big[\sqrt{-\det\boldsymbol{A}_{+}}+\sqrt{-\det\boldsymbol{A}_{-}}\Big],
 \end{equation}
where
\begin{equation}
\label{eq:Araddef}
\boldsymbol{A}_{(\pm)MN}=g_{MN}+ w(\lambda)(F^{V}_{MN}\pm F^{A}_{MN}),
\end{equation}
and 
\begin{align}
F^{V}_{MN}&=\partial_{M}V_{N}-\partial_{N}V_{M},\nonumber
\\F^{A}_{MN}&=\partial_{M}A_{N}-\partial_{N}A_{M}.
\end{align}

CP-odd effects are then modelled by the addition of the term which combines contributions both from the gluonic and flavor sectors \cite{Casero:2007ae,Arean:2013tja,Arean:2016hcs}
\begin{equation}
S_{a}=-\frac{M_{p}^{3}N_{c}^{2}}{2}\int d^{5}x\sqrt{-g}Z(\lambda)(\partial_{M}\mathfrak{a}-2xA_{M})^{2}\ ,\label{eq:actioncpodd}
\end{equation}
where the axion, $\mathfrak{a}$ is dual to the term  $\epsilon^{\mu\nu\rho\sigma}\Tr(G_{\mu\nu}G_{\rho\sigma})$, sourcing the $\theta$-angle. Here $\epsilon^{\mu\nu\rho\sigma}$ is the Levi-Civita symbol. This term is essential to our analysis because the mass term for $A_{M}$ present in the above action is necessary to include gluonic contributions to the model of anomalous transport~\cite{Jimenez-Alba:2014iia,Gallegos:2018ozs}.

The remaining piece, which turns out to be important for the analysis of the anomalous conductivites, is the Chern-Simons term  defined as
\begin{equation}
S_{\text{CS}}=-\frac{M_{p}^{3}N_{c}^{2}x}{2}\int d^{5}x\tilde{\epsilon}^{MNPQR}A_{M}(3\kappa F^{V}_{NP}F^{V}_{QR}+\gamma F^{A}_{NP}F^{A}_{QR}),
\end{equation}
where $\tilde{\epsilon}^{MNPQR}$ is the Levi-Civita symbol and $\kappa, \gamma$ are constant coefficients. Actually, the standard expressions for the Chern-Simons term in the literature~\cite{Witten:1998qj,Casero:2007ae} have fixed values for $\kappa$ and $\gamma$: 
\be
 \label{eq:kappaval}
 \kappa = \gamma = \frac{1}{24\pi^2M_p^3} \ .
\ee
This is because they are determined by the chiral flavor anomalies of QCD. We however choose to keep the coefficients unfixed for the moment for generality. The full Chern-Simons term, including effects from chiral symmetry breaking, has been analyzed recently in~\cite{Jarvinen:2022mys}.

Putting everything together, the holographic action to be studied is\footnote{In principle one should also add the Gibbons-Hawking and renormalization counterterms as this action is UV divergent, but these will not play any role in the analysis of this article.}
\begin{equation}
S=S_{g}+S_{f}+S_{a}+S_{\text{CS}}.
\end{equation}
As the full equations of motion are quite cumbersome and provide no physical insight before an Ansatz (to be presented in Sec.~\ref{subsec:background}) is assumed, they are listed in the App.~\ref{app:eoms}.

\subsection{Choice of the potentials}\label{subsec:choice_pot}

In order for the model to mimic the physics of QCD precisely, the potentials $V_g(\l)$, $V_f(\l)$, $w(\l)$, and $Z(\l)$ need to be chosen carefully. The asymptotics of the three first potentials ($V_g$, $V_f$, and $w$) both at weak and strong coupling are constrained by requiring that the model has fully regular solutions with confinement, discrete spectrum, correct mass gaps at finite quark mass, reasonable phase diagram with confined phase extending to finite chemical potential, asymptotically AdS solutions, and asymptotic freedom~\cite{Gursoy:2007cb,Gursoy:2007er,Jarvinen:2011qe,Arean:2012mq,Arean:2013tja,Jarvinen:2015ofa,Ishii:2019gta}. 

The details at intermediate coupling can then be fitted to either lattice data for QCD thermodynamics~\cite{Gursoy:2009jd,Panero:2009tv,Jokela:2018ers}, experimental meson masses~\cite{Amorim:2021gat} or to both~\cite{Jarvinen:2022gcc}. Since we are focusing on properties of the model at finite temperature in this article, we choose potentials which have been carefully fitted to lattice data for both Yang-Mills theory and full QCD in~\cite{Jokela:2018ers,Ishii:2019gta}. This procedure produces a family of models, which depends essentially on a single parameter, which cannot be determined by fitting. From this family, we use here the set 7a, which can be seen to be a conservative intermediate choice.  This set has been shown to also produce predictions at high density that are in good agreement with neutron star observations~\cite{Ecker:2019xrw,Jokela:2020piw,Jokela:2021vwy,Demircik:2021zll,Tootle:2022pvd}. See App.~\ref{app:potential} for explicit expressions for the potentials.

The potential $Z(\l)$, which appears in the $S_a$ term of the holographic action controls the physics of the axial anomaly -- it is also constrained by QCD data. Since this potential readily appears in the model for pure Yang-Mills ($x=0$), it is natural to fit it to the lattice data for related observables, including the topological susceptibility and the pseudoscalar glueball masses~\cite{Gursoy:2009jd}. This function also controls the Chern-Simons diffusion rate~\cite{Gursoy:2012bt,Bigazzi:2018ulg}, which measures the rate of change in the topological Chern-Simons number in Yang-Mills theory. 

We expect that the results for the anomalous conductivities in this article are more sensitive to the choice of $Z(\l)$ than the other potentials, which are quite strictly constrained by the fit to lattice data for the thermodynamics of QCD. Therefore, we analyze the effect of the choice of $Z(\l)$ on the results of the conductivities, keeping the other potentials fixed. We consider three choices:
\begin{enumerate}
    \item $Z(\l)$ with constant UV asymptotics,  i.e., $Z(\l) \sim \mathrm{const.}$ as $\l \to 0$.
    \item $Z(\l)$ with linear UV asymptotics,  i.e., $Z(\l) \sim \l$ as $\l \to 0$.
    \item $Z(\l)$ with quadratic UV asymptotics,  i.e., $Z(\l) \sim \l^2$ as $\l \to 0$.
 \end{enumerate}
In all these scenarios, we choose the IR asymptotics $Z(\l) \sim \l^4$ as $\l \to \infty $. With this choice, the spectra of the scalar and pseudoscalar glueballs have similar ``universal'' asymptotics at high excitation numbers~\cite{Gursoy:2007er}.

These three different choices for $Z(\l)$ near the boundary are motivated as follows. The constant choice was suggested in~\cite{Gursoy:2007er,Gursoy:2012bt} as the most general allowed term in analogy with effective theory. However, it turns out that this choice leads to a peculiar finite anomalous dimension for the axial chemical potential even in the free field theory limit $\l \to 0$~\cite{Arean:2016hcs,Weitz}. Moreover the Chern-Simons diffusion rate at high temperatures is higher with this choice~\cite{Gursoy:2012bt} as compared to the results from perturbative analysis~\cite{Arnold:1996dy,Bodeker:1998hm,Bodeker:1999gx,Moore:2010jd}. The linear choice for the asymptotics of $Z(\l)$ leads instead to an anomalous dimension which vanishes linearly in the coupling in the weak coupling limit. The UV running of the axial chemical potential is therefore similar to what one obtains for the quark mass in QCD. Finally, quadratic asymptotics for $Z(\l)$ matches with results from string perturbation theory~\cite{Gursoy:2007cb} and also agrees with computations in top-down models~\cite{Son:2002sd,Craps:2012hd,Bigazzi:2018ulg}, even though this comparison is not necessarily applicable at weak coupling. This choice is also interesting because, unlike the other choices, the running of the axial chemical potential is suppressed in the weak coupling limit (see Appendix~\ref{sec:ax_asym}) which allows us to define it unambiguously.

We emphasize that matching the weak coupling behavior of the function $Z(\l)$ with QCD results only gives indicative results at best because the holographic duality (with classical gravity approximation) is not expected to work at weak coupling. However, since the behavior of $Z(\l)$ at strong coupling is quite strictly fixed by the universality of the glueball trajectories, the weak coupling behavior is the main source of uncertainty for our results. Therefore it makes sense to explore different choices. These choices also drastically change the behavior of the background near the  boundary. Despite this, as we will discuss below, the effects on the results will be somewhat limited. We fit all functions $Z(\l)$ that we use to lattice data~\cite{Lucini:2005vg,Athenodorou:2021qvs,Bennett:2022gdz} for the topological susceptibility and glueball masses. See App.~\ref{app:potential} for details.

\subsection{Background solutions}\label{subsec:background}

For our analysis, we will solve the gravity backgrounds (charged planar black holes) for different values of the temperature, quark number chemical potential, and axial chemical potential. Recall that we are working at exactly zero quark mass and that we focus on the high-temperature deconfined phase of QCD so that chiral symmetry is intact. For the metric, we write
\be \label{eq:metric}
 ds^2 = g_{MN}dx^M dx^N = \frac{q(a)^2}{f(a)} da^2 + e^{2a}\left(-f(a)dt^2 + d\mathbf{x}^2\right)\ ,
\ee
which is a convenient Ansatz for numerical solutions in this model~\cite{Alho:2012mh,Alho:2013hsa}. Notice that the holographic coordinate $a$ maps to the logarithm of the energy scale on the field theory side as one can see from the warp factor of the spacetime coordinates in Eq.~\eqref{eq:metric}. Apart from the metric and the dilaton $\l(a)$, we will turn on the temporal components of both the vectorial and axial gauge fields, which we denote simply as $V(a)$ and $A(a)$, respectively. We choose a gauge where the bulk axion as well as the radial components of the gauge fields ($V_a$ and $A_a$) vanish\footnote{Our Ansatz for the axial gauge field, including its fluctuations, will be such that $\partial_MA^M =0$. This removes the coupling to the axion, which can be then consistently set to zero.}. Therefore the Ansatz is
\be
 V_M(x^N) = \delta_M^t V(a) \ , \qquad A_M(x^N) = \delta_M^t A(a)\ .
\ee
Note that for much of the rest of this article, we will suppress the $a$-dependence of $V(a),\,A(a)$.

To summarize, the equations of motion are then given by the Einstein equations, scalar equation for $\l$ and the gauge field equations, which are given for this setup in Appendix~\ref{app:eoms}.

The background solutions can be parameterized in terms of three parameters. For the numerical solutions, it is convenient to first give these parameters at the horizon of the black hole. They can be taken to be the horizon value $\l_h$ of the dilaton and the derivatives of the gauge fields at the horizon\footnote{Instead of the horizon derivatives, we actually define ``charges'' at the horizon that can be mapped to the values of the derivatives. See App.~\ref{app:eoms}.}. After solving the background numerically, these parameters can be mapped to the physical parameters, i.e., the temperature and the chemical potentials. 

The source of the dilaton field defines an additional parameter, $\Lambda$. This is the UV scale of the model, which is related to the 't Hooft coupling through dimensional transmutation, just as in QCD. See App.~\ref{app:eoms} for the precise holographic definition of $\Lambda$. The overall energy scale is however a symmetry of the solutions, so that the final three parameters characterizing the background are the dimensionless ratios $T/\Lambda$, $\mu_V/\Lambda$, and $\mu_A/\Lambda$. The physical value of the UV scale, $\Lambda \approx 211$~MeV, is obtained by comparing to lattice data~\cite{Jokela:2018ers} in the same way as the potential parameters discussed in App.~\ref{app:potential}.

\section{Anomalous conductivities}\label{sec:conductivities}

The anomalous conductivities for a vectorial current are defined through the linear response of the vectorial (i.e., quark number) current,
\be \label{eq:conddefs}
\delta\langle \mathcal{J}_V^k\rangle = \sigma_{VV} B^k + \sigma_{VA} B_A^k +\sigma_{V\Omega} \omega^k\ ,
\ee
where $k$ indicates spatial indices, $B$ is the (vectorial) magnetic field, $B_A$ is the axial magnetic field and $\omega$ is the vorticity. The magnetic fields and the vorticity are assumed to be small perturbations whereas the conductivities $\sigma_{VV}$, $\sigma_{VA}$, and $\sigma_{V\Omega}$ are associated with the chiral magnetic effect (CME), chiral separation effect (CSE) and chiral vortical effect (CVE) respectively. 

We apply here the definition Eq.~\eqref{eq:conddefs} for the consistent (rather than covariant) vectorial current. This is convenient in our model because the consistent current is divergence free, whereas the covariant current is UV divergent and needs to be renormalized. In this case, the CME and the CVE are determined by the anomaly structure~\cite{Son:2009tf,Neiman:2010zi,Landsteiner:2011cp,Gursoy:2014boa,Gursoy:2014ela,Grozdanov:2016ala} but the CSE does receive radiative corrections\footnote{When interpreting these results, note that most of the literature reports conductivities defined by the covariant current.\label{foot:cov_con}} 
\cite{Jimenez-Alba:2014iia,Gallegos:2018ozs}. We will therefore focus here on the CSE, analyzing such effects.

We make use of the holographic dictionary to derive the one-point 
function\footnote{We do not consider the one point function $\mathcal{J}_{A}$ here as its computation requires us to solve the equations of motion in full. Moreover, we only need to know $\mathcal{J}_{V}$ in order to compute the CME, CSE and CVE.}
\begin{align}
\langle\mathcal{J}_{V}^{\mu}\rangle&=\frac{1}{V_4}\frac{\delta S}{\delta V_\mu^\mathrm{bdry}} = \lim\limits_{a\rightarrow \infty} \frac{\partial \mathcal{L}}{\partial\left( \partial_a V_\mu\right) } = & \\
&= \lim\limits_{a\rightarrow \infty}\bigg\{\frac{M_{p}^{3}N_{c}^{2}xV_{f}(\lambda)w(\lambda)}{4}\sum_{s=+,-}\sqrt{-\det\boldsymbol{A}_{s}}\Big[\left(\boldsymbol{A}_{s}^{-1}\right)^{\mu a}-\left(\boldsymbol{A}_{s}^{-1}\right)^{a\mu}\Big]+\nonumber\\
& \ \ \ \ +12M_{p}^{3}N_{c}^{2}x\kappa{\epsilon}^{\sigma\nu\rho\mu}A_{\sigma}\partial_{\nu}V_{\rho}\bigg\}\ ,\label{eq:conscurrent}
\end{align}
where $V_4$ is the volume of the space-time, $\boldsymbol{A}_\pm$ were given in Eq.~\eqref{eq:Araddef}, the index $a$ denotes the holographic direction and $V_\mu^\mathrm{bdry} = \lim_{a\to \infty} V_\mu(a)$. Notice that this is the generic expression which holds for any background assuming that the current is independent of space-time coordinates. That is, we did not insert the Ansatz for the background discussed above, but we did assume the standard radial gauge for the bulk gauge fields.
We identify this one-point function as the consistent vectorial current. Observing the response of this current to a perturbation will enable us to calculate the anomalous coefficients in Eq.~\eqref{eq:conddefs}. 

Adding small external magnetic fields and vorticity amounts to adding infinitesimal sources to the field strength tensors of the gauge fields and the shear components of the metric. Constant magnetic fields are obtained by turning on perturbations of the gauge fields that are linear in one of the spatial coordinates, which we choose to be the $z$-coordinate. That is, we consider the following perturbations\footnote{To have a fully consistent system of fluctuations, one also needs to add terms that are independent of $z$. These terms however will not contribute to the conductivities.}
\begin{align}
 \delta V_k(x^M) &= z \delta \hat V_k(a) \ , \qquad  \delta A_k(x^M) = z \delta \hat A_k(a) \ , \\
 \delta g_{tk}(x^M) &=  e^{2a}h_{tk}(x^M) = e^{2a} z \hat h_k(a) &
\end{align}
on top of the background defined in subsection~\ref{subsec:background}. Here, only the indices $k=x,y$ are relevant and we again suppress the explicit a-dependence in much of what follows.

This perturbation gives rise to the following vorticity and (vectorial) magnetic field:
\begin{align}
 \label{eq:vorticitydef}
 \omega^i &= \lim_{a\to\infty}\epsilon^{ijk}\partial_j h_{tk} = - \epsilon^{zij} \lim_{a\to\infty} \hat h_j(a)\ , &\\
 B^i - \mu_V \omega^i &= \frac{1}{2}\lim_{a\to\infty}\epsilon^{ijk}\delta F^V_{jk} = -\epsilon^{zij}\lim_{a\to\infty}\delta \hat V_j(a)\ . &
\end{align}
$B^i$ is the magnetic field in the rest frame of the fluid, corrected by the field $\mu_V \omega^i$ arising from the infinitesimal motion of the fluid.   
The definition of the axial magnetic field is otherwise similar to the vectorial field albeit with the additional complication that the field runs close to the boundary, which needs to be taken into account. See App.~\ref{sec:ax_asym} for details. 

The fluctuation equations for the fields $\delta \hat V_k$, $\delta \hat A_k$, and  $\hat h_k$ with $k=x$ and $k=y$ are identical. In addition, the equations for different values of the index are decoupled. Therefore we henceforth omit this index and write

\begin{subequations}
\label{eq:fluceqs}
\begin{align}
\label{eq:Lfluct}
 x  e^{2a} q Z(\lambda ) \delta \hat A&= \frac{d}{d a}\left[ \frac{f e^{2a} V_f(\lambda ) w(\lambda )^2}{4 q R_+}\left(\delta \hat V'+\delta \hat A'\right)\right] +\frac{\left(V'+A'\right) e^{2a} V_f(\lambda ) w(\lambda )^2}{4 q R_+} \hat h'  \\
\label{eq:Rfluct}
  -x  e^{2a} q Z(\lambda )\delta \hat A &= \frac{d}{d a}\left[ \frac{f e^{2a} V_f(\lambda ) w(\lambda )^2}{4 q R_-}\left(\delta \hat V'-\delta \hat A'\right)\right] +\frac{\left(V'-A'\right) e^{2a} V_f(\lambda ) w(\lambda )^2}{4 q R_-} \hat h' \\
  C_f 
  &=\frac{\left(V'+A'\right) x e^{2a} V_f(\lambda ) w(\lambda )^2}{2 q R_+}\left(\delta \hat V+\delta \hat A\right)+ &\nonumber\\
  &\ \ \ +\frac{\left(V'-A'\right) x e^{2a} V_f(\lambda ) w(\lambda )^2}{2 q R_-} \left(\delta \hat V-\delta \hat A\right)+\frac{e^{4a} }{q}\hat h'  &
  \label{eq:fluctEinstein}
\end{align}
\end{subequations}
where $C_f$ is a constant of integration and 
\be \label{eq:Rpmdef}
 R_\pm = \sqrt{1-\frac{e^{-2 a} w(\lambda )^2 \left(A'\pm V'\right)^2}{q^2}} \ .
\ee
A general solution to this system has five other 
constants of integration, which can be associated with the boundary conditions at the horizon. Horizon regularity however fixes two of these constants, which can be thought of as derivatives of the gauge fields at the horizon. An additional special condition for time-independent fluctuations is that the fluctuation of the metric, $\hat h$, must vanish at the horizon for the fluctuated scalar curvature $R$ to remain finite~\cite{Gursoy:2014boa}, so that 
the fluctuated action is normalizable.

The remaining three constants then map to the three observables, i.e., the conductivities from Eq.~\eqref{eq:conddefs}. 

To compute the conductivities we therefore need to find three regular solutions to the fluctuations. Two such solutions can however be found immediately. For the first solution, $\delta \hat V$ is constant:
\be
\label{eq:firstsol}
\delta \hat V = |B| \ , \qquad \delta \hat A = 0\ , \qquad \hat h = 0\ .
\ee
To check this solution, note that the coefficient of $\delta \hat V$ in Eq.~\eqref{eq:fluctEinstein} is also a constant: it is denoted as $c_V$ in Eq.~\eqref{eq:Vint} in Appendix~\ref{app:eoms}. Therefore, Eq.~\eqref{eq:firstsol} solves the fluctuation equations for a specific choice of $C_f$.
The second solution is
\be
\label{eq:secondsol}
\delta \hat V = -|\omega| V \ , \qquad \delta \hat A = -|\omega| A\ , \qquad \hat h = |\omega| f\  .
\ee
This solution corresponds to an infinitesimal boost of the background. It can be checked by inserting it in the fluctuation equations and by using the background equations of motion Eqs.~\eqref{eq:pluseom},~\eqref{eq:minuseom} and~\eqref{eq:fint}. The remaining solution (for which $\delta \hat A$ is non-vanishing at the horizon) can then be constructed numerically.

Let us then compute the response of the current Eq.~\eqref{eq:conscurrent} to the small perturbations. First we note that, because the action only depends on $V_\mu$ through the field strength tensor, we can immediately use the equation of motion for $V_\mu$ to write (assuming again radial gauge)
\be
 \delta \langle\mathcal{J}_{V}^{\mu}\rangle = \delta \int_{a_h}^\infty da\ \partial_\nu \frac{\partial \mathcal{L}}{\partial\left(\partial_\nu V_\mu\right)} \ ,
\ee
where $a_h$ denotes the location of the horizon and we assumed that the perturbation of the IR term (the charge of the black hole) vanishes. For the perturbation we consider here, only the Chern-Simons action contributes, leaving us with
\begin{align}
 \delta \langle\mathcal{J}_{V}^{i}\rangle &= 12 M_p^3 N_c^2 x \kappa \epsilon^{zij} \int_{a_h}^\infty da \left(A\ \partial_z \partial_a \delta V_j -\partial_a V\ \partial_z  \delta A_j\right) & \\
 &= \frac{N_cN_f}{2\pi^2}\,\epsilon^{zij} \int_{a_h}^\infty da \left(A \  \delta \hat V_j' - V'\   \delta \hat A_j\right)  \ , &
 \label{eq:currentfinal}
\end{align}
where we inserted the value of $\kappa$ from Eq.~\eqref{eq:kappaval}.

Notice that the current vanishes for both the solutions Eqs.~\eqref{eq:firstsol} and~\eqref{eq:secondsol}. These solutions turn on nonzero vectorial magnetic field and vorticity respectively.
Consequently, the chiral magnetic and vortical effects vanish, 
\be \label{eq:zeroconds}
\sigma_{VV} =0 = \sigma_{V\Omega} \ .
\ee

The third conductivity $\sigma_{VA}$ can be computed numerically. We can however analyze some of its properties by studying the discrete symmetries of the action. The fluctuations and equations of motion are unchanged under two discrete symmetries: the simultaneous change of sign of the vectorial (axial) gauge field and its fluctuation. The current in Eq.~\eqref{eq:currentfinal} is odd under both symmetries, whereas the axial magnetic field is even under the vectorial symmetry but odd under the axial symmetry. Therefore $\sigma_{VA}$ is odd (even) under the vectorial (axial) symmetry. Moreover, the vectorial (axial) symmetry changes the sign of the vectorial (axial) chemical potential. Therefore $\sigma_{VA}$ is an odd function of $\mu_V$ and even function of $\mu_A$. In particular, $\sigma_{VA}$ vanishes for $\mu_V=0$.

Finally, let us comment on this result in the limit of large temperature. In this weak coupling limit: the black hole becomes large and the dilaton $\lambda \sim 1/a$~\cite{Gursoy:2007cb,Gursoy:2007er} is small for all values of the holographic coordinate $a$. If $Z(\l)$ vanishes sufficiently fast as $\l \to 0$, it can be set to zero in Eqs.~\eqref{eq:Lfluct} and~\eqref{eq:Rfluct}. Then the third solution of the fluctuation equations also becomes simple:
\be
\label{eq:thirdsol}
\delta \hat V \approx 0 \ , \qquad \delta \hat A \approx |B_A|\ , \qquad \hat h \approx 0\ .
\ee
As one can check, this only holds for choice 3 of the function $Z(\l)$ with quadratic UV asymptotics; the analysis of Appendix~\ref{sec:ax_asym} excludes such a solution for choices 1 and 2.

For the choice 3 we find
\be
\delta \langle\mathcal{J}_{V}^{i}\rangle \approx -\frac{N_cN_f}{2\pi^2}\,\epsilon^{zij}    \delta \hat A_j \int_{a_h}^\infty da  \,V' = \frac{N_cN_f}{2\pi^2} \mu_V B_A^i \ , 
\ee
so that we recover the universal result~\cite{Son:2009tf,Neiman:2010zi,Landsteiner:2011cp,Gursoy:2014boa,Gursoy:2014ela,Grozdanov:2016ala}\footnote{To reiterate (see footnote~\ref{foot:cov_con}), some of these references present the conductivities obtained from the covariant current as opposed to the consistent current, as we do here.}
\be
 \sigma^{U}_{VV} = 0 \ , \qquad \sigma^{U}_{VA} = \frac{N_cN_f}{2\pi^2}\mu_V \ , \qquad \sigma^{U}_{V\Omega} = 0 \ , \qquad (T \to \infty)\ ,
 \label{eq:univconds}
\ee
This universal result is also found in the free theory~\cite{Metlitski:2005pr,Brandt:2023wgf} so it is indeed the expected high-temperature behavior in QCD, which becomes weakly coupled as the temperature grows.

\section{CSE results}\label{sec:results}
Upon solving the fluctuation equations Eq.~\eqref{eq:fluceqs} on top of the background equations Eq.~\eqref{eq:bgeqs}, we compute the fluctuation of the vector current, Eq.~\eqref{eq:currentfinal} from which we can obtain the various anomalous conductivities. While the conductivity associated with the CME and CVE are zero and therefore do not deviate from their universal values (see Eqs.~\eqref{eq:zeroconds} and~\eqref{eq:univconds}), the CSE does indeed receive radiative corrections, which we now proceed to discuss. 

Throughout our discussion, we generally refrain from commenting on the absolute size of the radiative corrections that we find, due to the arbitrary normalization associated with the axial magnetic field, $B_{A}$ for choices $1$ and $2$ of the $Z(\lambda)$ potential (see Appendix~\ref{sec:ax_asym}). For choice $3$, this arbitrariness is however absent, allowing us to take seriously the corrections' size. 

The CSE has recently been computed on the lattice~\cite{Brandt:2023wgf} by analyzing the effect of the magnetic field on the axial current. Said work computes the CSE as a function of temperature, finding it to be suppressed with respect to the universal result in the confining regime. A steep rise in the conductivity around the transition temperature then leads to a marginal enhancement with respect to the universal result at higher temperatures ($T\gtrsim 400~ \text{MeV}$).

\begin{figure}[h]
\centering
  \includegraphics[width=1\textwidth]{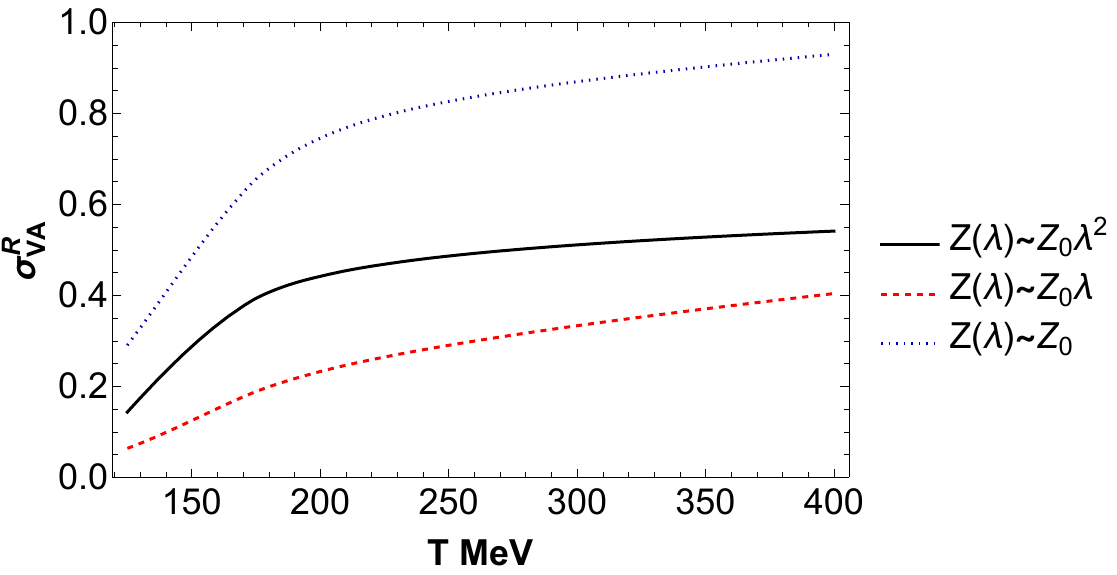}
\caption{The ratio $\sigma^{R}_{VA}\equiv\frac{\sigma_{VA}}{\sigma^{U}_{VA}}$ plotted as a function of temperature for different choices of $Z(\lambda)$ potential with $\mu_{V}=1\,\text{keV}$ and zero $\mu_{A}$.}
\label{fig:diffchoices_diffT}
\end{figure}

\begin{figure}[h]
\centering
  \includegraphics[width=0.9\textwidth]{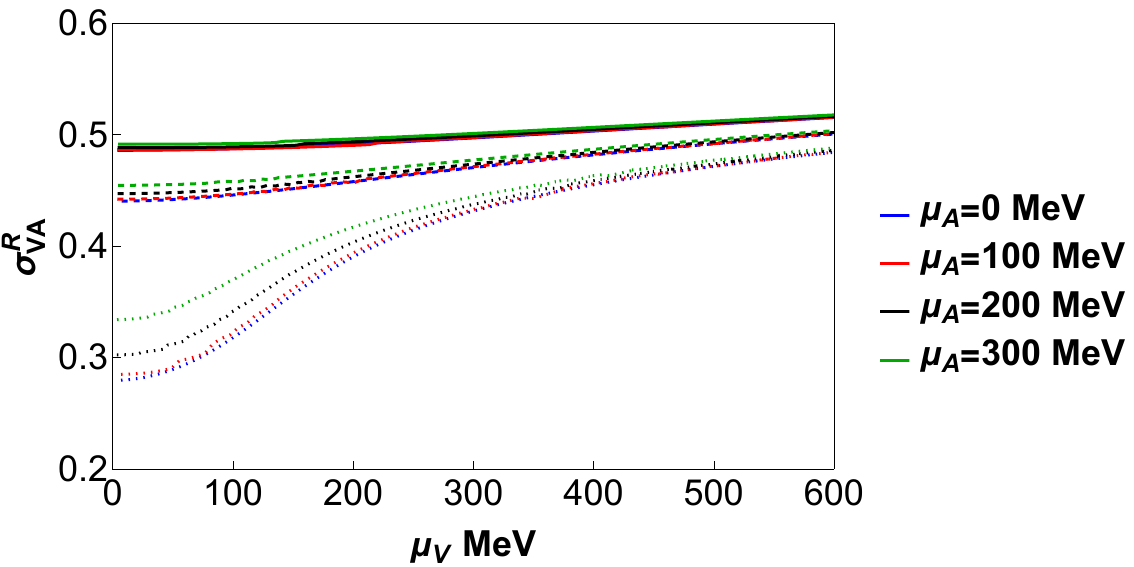}
\caption{The ratio $\sigma^{R}_{VA}\equiv\frac{\sigma_{VA}}{\sigma^{U}_{VA}}$ 
plotted as a function of vector chemical potential for choice 3 of the $Z(\lambda)$ potential. Solid lines correspond to $T=250\,\text{MeV}$, dashed lines to $T=200\,\text{MeV}$ and dotted lines to $T=150\,\text{MeV}$. The definition of $\mu_{A}$ varies according to the choice of $Z(\l)$ -- see App.~\ref{sec:ax_asym}.}
\label{fig:cse_2}
\end{figure}

In Fig.~\ref{fig:diffchoices_diffT}, we show our results for the CSE as a function of temperature, for small vector chemical potential and zero axial chemical potential, so as to emulate the lattice setup. We observe similar behaviour to that found in~\cite{Brandt:2023wgf}, at least qualitatively, for all three choices of the $Z(\lambda)$ potential; a noticeable change in slope is recovered around the critical temperature, giving rise to a knee-like feature.  
The results for none of the choices of potentials seem to agree with the universal or free theory results at large temperatures. This is not surprising for choices $1$ and $2$ since, as we remarked above, for these choices, the conductivity does not approach the free theory value in the high-temperature limit, since its normalization is arbitrary. For choice $3$ however the free theory value will be eventually obtained as $T \to \infty$, albeit with a very gradual convergence.

It is perhaps unsurprising that the high-temperature result in V-QCD does not automatically agree with the free field limit 
given that V-QCD is by definition a strongly coupled model, fitted to reproduce low-temperature features of QCD. Indeed, in general, the gauge/gravity duality is not expected to work at weak coupling. It would nevertheless be interesting to see how the addition of quark masses via the tachyon field in our setup would impact the behaviour of the conductivity around the critical temperature. We will come back to discuss this in the Conclusion. 

One might also be worried by the fact that even for potentials 3, for which the normalization of the CSE conductivity is well-defined, the results in Fig.~\ref{fig:diffchoices_diffT} differ from the lattice results~\cite{Brandt:2023wgf} by a largish factor at low temperatures, i.e., in the strongly coupled region. However, we think that this disagreement is also a weak-coupling effect: The integral in the expression for the current in~\eqref{eq:currentfinal} is dominated by the near-horizon region and therefore essentially only depends on 
the values of the fluctuation of axial gauge field $\delta \hat A$ 
in this region. The 
magnetic field $B_A$ is however evaluated at the boundary, thus probing the value of $\delta \hat A$  
in the ultimate weak-coupling limit. 

The flow of $\delta\hat A$ between the horizon and boundary region contains the contribution from the region which is described\footnote{One can check numerically that the near-boundary expansion is a good approximation for $a \gtrsim 10$.} by the near-boundary expansion given in Appendix~\ref{sec:ax_asym} and independent of the temperature.
This flow gives rise to a constant contribution in the CSE conductivity, i.e., a weak-coupling contribution that is not expected to be reliably captured by the holographic model.

\begin{figure}
\centering
  \includegraphics[width=0.9\textwidth]{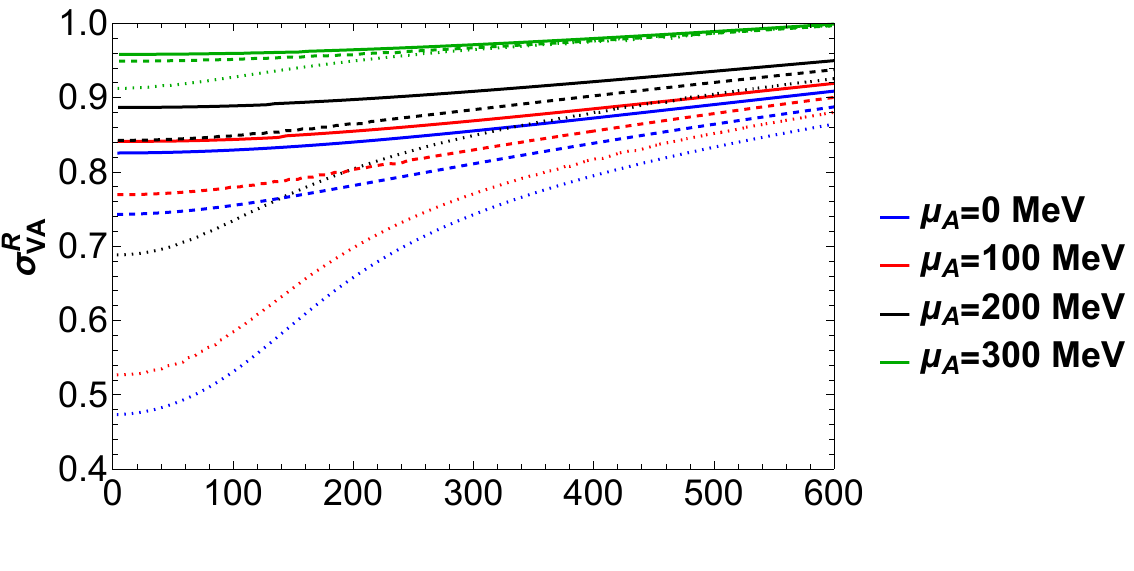}
  \newline
  \includegraphics[width=0.9\textwidth]{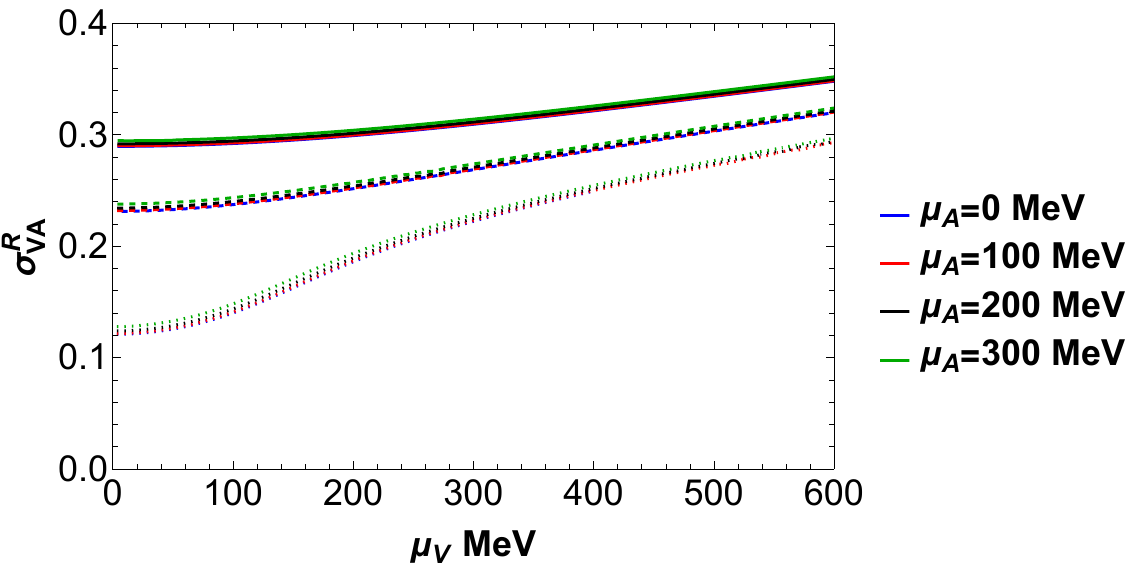}
\caption{The ratio $\sigma^{R}_{VA}\equiv\frac{\sigma_{VA}}{\sigma^{U}_{VA}}$ plotted as a function of vector chemical potential. From top to bottom, figures are for choices 1 and 2 of the $Z(\lambda)$ potential. Solid lines correspond to $T=250\,\text{MeV}$, dashed lines to $T=200\,\text{MeV}$ and dotted lines to $T=150\,\text{MeV}$. The definition of $\mu_{A}$ varies according to the choice of $Z(\l)$ -- see App.~\ref{sec:ax_asym}.}
\label{fig:cse}
\end{figure}

In Figs.~\ref{fig:cse_2} and~\ref{fig:cse}, we show our results for the CSE computed as a function of the vectorial chemical potential, for fixed values of the axial chemical potential. The former shows that for choice $3$ of the $Z(\lambda)$ potential, the CSE 
is suppressed with respect to the universal result across a range of vectorial and axial chemical potentials in the strong-coupling regime. The ratio to the universal conductivity is an increasing function of all variables $T$, $\mu_V$, and $\mu_A$, but the dependence on each of the variables is somewhat different.
The conductivity curves widen non-linearly as a function of the axial chemical potential at fixed temperature -- this is particularly noticeable for the $T=150~\text{MeV}$ curves at low vector chemical potential. Overall, the corrections clearly exhibit a stronger dependence on the temperature, which nevertheless begins to disappear for larger values of $\mu_{V}$, as the various curves appear to asymptotically converge. 

Moving on to Fig.~\ref{fig:cse}, we see in general the same behaviour for the CSE for choices $1$ and $2$ of the $Z(\lambda)$ potential. Even though we cannot deduce a suppression or enhancement with respect to the universal result, we note an apparent weakening of the temperature (and axial chemical potential) dependence for large values of the vector chemical potential. One feature worth commenting on for choice $1$ of the $Z(\lambda)$ potential is the rather strong dependence on the axial chemical potential; the nonlinear dependence on $\mu_{A}$ overtakes the temperature dependence in this case. 
The large difference between choices $1$ and $2$ is apparently due to differences in the normalization of $\mu_A$ between the potentials\footnote{A direct way to estimate the effect of the chemical potential on the background is to evaluate the charges $n_A$ defined in Appendix~\ref{app:eoms}. We find that for fixed $\mu_A$ the charges for choices $1$ and $3$ are roughly of the same size, while the charge for the choice $2$ is significantly smaller. This is in rough agreement with the observed dependence on $\mu_A$ in Figs.~\ref{fig:cse_2} and~\ref{fig:cse}.}. Recall that there is no obvious way to fix these normalizations (see Appendix~\ref{sec:ax_asym}).
Notice also that the coefficients $c_i$ defined in Appendix~\ref{app:z_anal} are much larger for the choice 1 of the $Z(\lambda)$ potential -- we remind the reader that these coefficients are fixed in order to obtain a good agreement with lattice values of the pseudoscalar glueball spectrum.

\section{Conclusions}\label{sec:conclusions}

In this article, we have discussed anomalous conductivities in a bottom-up holographic model, V-QCD. This model has been determined through a careful comparison to QCD data, in particular lattice data for the thermodynamics of QCD at nonzero temperatures and small densities.

We considered a simple method, where we turned on constant (vectorial and axial) magnetic fields as well as vorticity and read off the anomalous conductivities for the response of the (consistent) vectorial current. Our results agree with the more general derivations from the literature: in particular, the CME and CVE vanish in our setup (in the conventions which use the consistent vectorial current to define the anomalous conductivities). 

Therefore, the only nonzero conductivity is that which is related to the CSE. The value of this conductivity is not fixed by the anomalies but receives radiative corrections in QCD. We computed these corrections using different variants of the holographic V-QCD model, and found results in the limit of small density (see Fig.~\ref{fig:diffchoices_diffT}), which qualitatively agree with the recent findings from lattice QCD simulations~\cite{Brandt:2023wgf}. We gave predictions for the dependence on vectorial chemical potential in Figs.~\ref{fig:cse_2} and~\ref{fig:cse}.

Our $\mu_{V}$-dependent evaluation of the CSE (Figs.~\ref{fig:cse_2} and~\ref{fig:cse}) could be plausibly relevant for heavy-ion experiments at the LHC~\cite{Wang:2023xhn} and at RHIC~\cite{STAR:2021mii}. In particular, the recent isobar run at RHIC has been devoted to finding conclusive evidence of the CME\footnote{Observation of such a signal has however so far remained elusive.}. Our results could also be relevant for the search of the CMW\footnote{In more detail, it has been argued that it should be possible to identify a CMW signal, even in the absence of the CME and CSE observations. This stems from the fact that the CMW can occur in the presence of zero average background axial and vector charge density (which are needed to be non-zero to give rise to the CME and CSE respectively). Indeed, the CMW can be generated in chiral matter by density fluctuations as long as there is an external magnetic field.}, which, while to a lesser extent than the CME, is also the subject of current experimental investigation~\cite{CMS:2017pah,STAR:2022zpv}.

Our analysis could be extended in various ways. The most obvious extension would be to analyze the axial current in addition to the vectorial current of Eq.~\eqref{eq:conscurrent}. This requires carrying out holographic renormalization, and would give information about two additional conductivities ($\sigma_{AA}$ and $\sigma_{A\Omega}$ in notation analogous to Eq.~\eqref{eq:conddefs}). Another possible avenue would be to check whether our analysis and results can be connected to the new method of analyzing holographic transport suggested in~\cite{Demircik:2023lsn,Demircik:2024bxd}. It would also be interesting to study dynamical setups, which mimic more closely the evolution of the plasma after heavy-ion collisions, for example following recent studies on somewhat simpler St\"uckelberg models~\cite{Grieninger:2023myf,Grieninger:2023wuq}.

Note also that  we did not include quark masses -- all quark flavors were assumed be massless and identical. While the light quark masses are expected to be small perturbations to the current computation, the strange quark mass might have a significant effect on the results. Checking this properly requires adding the strange quark in the V-QCD model or in some simpler holographic model fitted to lattice data. In V-QCD, this extension requires adding an additional scalar field, the tachyon, in the DBI action Eq.~\eqref{eq:DBI} (see~\cite{Jarvinen:2015ofa,CruzRojas:2024etx}). Moreover, as we did not include electric charges for the quarks, our magnetic fields couple to all quarks in the same way. Be that as it may, since we are only studying linear conductivities in these fields, adding charges will not change the results qualitatively. Related to this, we remark that the lattice study~\cite{Brandt:2023wgf} did analyze flavor dependence of the CSE and found qualitatively similar temperature dependence in basically all flavor channels. 

It may nevertheless be interesting to try to fit the flavor dependence of the lattice data more precisely. We expect that doing this accurately requires the model extension which contains the strange quark mass. It might also be possible to find a function $Z(\lambda)$ which agrees with the fast approach to the universal or free quark result with increasing temperature observed on the lattice. As we argued, this requires that the function vanishes fast enough as weak coupling, as it controls the renormalization group flow of the axial chemical potential and magnetic field, which should be suppressed for the universal result to hold. We speculate that choosing asymptotics that vanish even faster than the quadratic asymptotic which we studied here might suppress the flow further and lead to faster convergence towards the universal result.

\acknowledgments

We thank J.~Ghiglieri, S.~Grieninger, U.~G\"ursoy, N.~Jokela, N.~Poovuttikul for discussions. M.~J. has been supported by an appointment to the JRG Program at the APCTP through the Science and Technology Promotion Fund and Lottery Fund of the Korean Government. M.~J. has also been supported by the Korean Local Governments -- Gyeong\-sang\-buk-do Province and Pohang City -- and by the National Research Foundation of Korea (NRF) funded by the Korean government (MSIT) (grant number 2021R1A2C1010834). 

\appendix
\renewcommand{\thesection}{\Alph{section}}
\renewcommand{\thesubsection}{\Alph{section}.\arabic{subsection}}
\renewcommand{\theequation}{\Alph{section}.\arabic{equation}}

\section{Potential definitions}
\label{app:potential}

We pack into this appendix the definitions of the various V-QCD potentials $V_{g}(\lambda), V_{f}(\lambda)$ and $w(\lambda)$ as well as the potential $Z(\lambda)$, all of which enter into our holographic model, described in Sec.~\ref{sec:vqcd}. 

We start with the gluonic potential
\begin{equation}
V_{g}(\lambda)=12\Bigg[1+V_{1}\lambda+\frac{V_{2}\lambda^{2}}{1+\lambda/\lambda_{0}}+V_{\text{IR}}e^{-\lambda_{0}/\lambda}(\lambda/\lambda_{0})^{4/3}\sqrt{\log(1+\lambda/\lambda_{0})}\Bigg].
\end{equation}
The coefficients $V_{1}$ and $V_{2}$ are fixed by requiring the UV RG flow of the 't Hooft coupling to be the same as in QCD up to two-loop order. The IR coefficients are then chosen to reproduce qualitative features of QCD such as confinement, magnetic charge screening and linear glueball trajectories through a comparison with lattice data \cite{Panero:2009tv}.
In addition, the potentials for the flavor sector are
\begin{align}
V_{f}(\lambda)&=W_{0}+W_{1}\lambda+\frac{W_{2}\lambda^{2}}{1+\lambda/\tilde{\lambda}_{0}}+12 W_{\text{IR}}e^{-\tilde{\lambda}_{0}/\lambda}(\lambda/\tilde{\lambda}_{0})^{2},
\\\frac{1}{w(\lambda)}&=w_{0}\Bigg[1+\bar{w}_{0}e^{\hat{\lambda}_{0}/\lambda}\frac{(\lambda/\hat{\lambda}_{0})^{4/3}}{\log(1+\lambda/\hat{\lambda}_{0})}\Bigg].
\end{align}
The UV coefficients of $V_f$ are determined by comparison with the flavor dependence of the QCD beta function, while the powers appearing in the IR dominant terms are chosen to reproduce some of the prominent features of the QCD phase diagram, such as chiral symmetry breaking and linear meson trajectories. The coefficient $W_0$ cannot however be determined by comparing to the beta function and remains as a free parameter.

Matching with perturbative QCD gives for the UV coefficients~\cite{Gursoy:2007cb,Jarvinen:2011qe}
\begin{align}
V_{1}&=\frac{11}{27\pi^{2}}&&,&&V_{2}=\frac{4619}{46656\pi^{4}}&,&&\nonumber\\
W_{1}&=\frac{8+3W_{0}}{9\pi^{2}}&,&&W_{2}&=\frac{6488+999W_{0}}{15552\pi^{2}}\ .
\end{align}

Furthermore, the IR coefficients and $w_0$ are fitted to lattice data~\cite{Gursoy:2009jd,Jokela:2018ers,Ishii:2019gta}. We choose the fit 7a in these references, which means the choice
\begin{align}
V_{\text{IR}}&=2.05&,&&\lambda_{0}&=\frac{8\pi^{2}}{3}&,&& \nonumber\\
W_{0}&=2.5&,&&W_{\text{IR}}&=0.9 & ,&& \nonumber\\
\hat{\lambda}_0&=\frac{8\pi^2}{1.18}&,&&w_{0}&=1.28&,&&\bar{w}_0&=18\ ,
\end{align}
for the remaining parameters. Moreover, the lattice fit also fixes
\be
 \Lambda \approx 210.8\ \mathrm{MeV} \ , \qquad M_p \approx 0.1972 \ .
\ee

\subsection{$Z(\lambda)$ Analysis}\label{app:z_anal}
As has been highlighted in the main text, the inclusion of a mass term for the $A_M$ field (see Eq.~\eqref{eq:actioncpodd}) allows for the study of radiative corrections to the anomalous conductivities, which we present in Sec.~\ref{sec:results}. We devote the rest of this appendix to a discussion of the $Z(\lambda)$ potential, considering various choices and detailing our methods for fixing its coefficients. To start, let us consider the general Ansatz 
\begin{equation}
    Z(\lambda)=Z_0\bigg(c_0+c_1\Big(\frac{\lambda}{\tilde{\lambda}_0}\Big)+c_2\Big(\frac{\lambda}{\tilde{\lambda}_0}\Big)^2+c_4\Big(\frac{\lambda}{\tilde{\lambda}_0}\Big)^4\bigg),
\end{equation}
where $\tilde{\lambda}_0=8\pi^2$. Referring to the discussion in Sec.~\ref{subsec:choice_pot}:
\begin{itemize}
    \item Choice 1 leads to $c_0=1$ and $c_2=0$.
    \item Choice 2 leads to $c_0=c_2=0$ and $c_1=1$.
    \item Choice 3 leads to $c_0=c_1=0$ and $c_2=1$.
\end{itemize}
We then follow the procedure described in \cite{Gursoy:2007er,Gursoy:2012bt}, tweaking the remaining unfixed $c_i$ for each choice in such a way that the pseudo-scalar, lattice-calculated glueball spectra \cite{Athenodorou:2021qvs,Bennett:2022gdz} are best reproduced. 
 
After all of the $c_i$ are fixed, the coefficient $Z_0$ can then obtained \cite{Gursoy:2007er,Gursoy:2009jd,Gursoy:2012bt} from the topological susceptibility (in pure Yang-Mills theory), $\chi_{\text{YM}}$. In $a$ coordinates, it is computed from 
\begin{equation}    
    Z_0=-\frac{\chi_{\text{YM}}}{M_p^3}\int_{-\infty}^{\infty} da\frac{e^{-4a}q}{c_0+c_1\Big(\frac{\lambda_{\text{YM}}}{\tilde{\lambda}_0}\Big)+c_2\Big(\frac{\lambda_{\text{YM}}}{\tilde{\lambda}_0}\Big)^2+c_4\Big(\frac{\lambda_{\text{YM}}}{\tilde{\lambda}_0}\Big)^4},
    \label{eq:zeq_def}
\end{equation}
where $\lambda_{\text{YM}}$ is the dilaton 
field in the pure glue theory. In this instance, we use the lattice result 
$\chi_{\text{YM}}/\sigma^2 = 0.01937(136)$ \cite{Bennett:2022gdz}, with $\sigma$ the string tension. Instead of computing the string tension from the gravity side of the theory, we choose to work with the dimensionless ratio $\chi_{\text{YM}}/T_c^4$; the value of $T_c/\sqrt{\sigma}=0.5970(38)$ comes from \cite{Lucini:2005vg}, so that $\chi_{\text{YM}}/T_c^4\approx 0.152$. With this recipe in hand, we can fix all of the coefficients for each choice of the $Z(\lambda)$ potential. The various inputs are collected in Table.~\ref{table:z_coeffs}.
\begin{table}[h!]
	\centering
	\begin{tabular}{ |p{1.5cm}|p{2.5cm}|p{1.5cm}|p{0.5cm}|p{1cm}|p{0.5cm}|p{1.3cm}|  }
		\hline
		Choice & UV behaviour & $Z_0$ & $c_0$ & $c_1$ & $c_2$ & $c_4$\\
		\hline
		\hline
		1 & $\text{const.}$ & $0.443387$ & $1$ & $21.907$ & $0$ & $3$\\
		\hline
		2 & $\lambda$ & $9.04723$ & $0$ & $1$ & $0$ & $0.2224$\\
		\hline
		3 & $\lambda^2$ & $14.0949$ & $0$ & $0$ & $1$ & $0.68954$\\
		\hline
	\end{tabular}
	\caption{Results for different choices of $Z(\lambda)$ potential.  Coefficients are  defined according to Eq.~\eqref{eq:zeq_def}.}
	\label{table:z_coeffs}
\end{table}

A priori, it may not be clear why there are $3$ non-zero $c_i$ for the case of choice $1$, as there are only two non-zero $c_i$ for choice $2$ and $3$. Indeed, for choice $1$, it turns out\footnote{See App.~\ref{sec:ax_asym} for more detail.} that it is necessary to have non-zero $c_1$ in order to yield the desired UV asymptotics for the axial gauge field.

\section{Equations of motion and the background}\label{app:eoms}

The equations of motion for the background consist of the Einstein equations, the equation for the scalar dilaton field and the equations for the vectorial and axial gauge fields. As usual, this is not a fully independent system of equations, but for example the dilaton equation can be derived from the rest (also using the zero energy constraint included in the Einstein equations). We present here the equations only for the Ansatz we will use in this article, i.e., taking the metric of Eq.~\eqref{eq:metric}, assuming that the dilaton only depends on $a$ and writing for the gauge fields
\be
 V_M(x^N) = V(a)\delta^t_M \ , \qquad A_M(x^N) = A(a)\delta^t_M \ .
\ee
The equations of motion then reduce to the following system:
\begin{subequations}
\label{eq:bgeqs}
\begin{align}
\label{eq:lambdaeq}
\frac{\left(\lambda '\right)^2}{\lambda^2 }&=9+\frac{9}{4}\frac{f'}{f}+\frac{3}{8} \frac{q^2}{f} xV_f(\lambda ) \left(\frac{1}{R_+}+\frac{1}{R_-}\right)-\frac{3}{4}  \frac{q^2}{f} V_g(\lambda )- & \nonumber \\
&\ \ \ -\frac{3 x^2Z(\lambda ) e^{-2a} q^2}{2 f^2}A^2\ ,&\\
\label{eq:qeq}
\frac{q'}{q}&=4+\frac{f'}{f}+\frac{1}{6} \frac{q^2}{f} xV_f(\lambda ) \left(\frac{1}{R_+}+\frac{1}{R_-}\right) -\frac{1}{3} \frac{q^2}{f} V_g(\lambda ) \ ,&\\
\label{eq:pluseom}
 \frac{e^{2a} q xZ(\lambda )}{f} A &=  \frac{d}{da}\left[\frac{e^{2a} V_f(\lambda ) w(\lambda )^2 \left(V'+A'\right)}{4 q R_+}\right]\ , & \\
\label{eq:minuseom}
- \frac{e^{2a} q xZ(\lambda )}{f} A &=  \frac{d}{da}\left[\frac{e^{2a} V_f(\lambda ) w(\lambda )^2 \left(V'-A'\right)}{4 q R_-}\right]\ , & \\
f''&=\frac{f'q'}{q}-4f' 
+\frac{e^{-2a}xV_f(\lambda)w(\lambda)^2}{2}\left[\frac{(V'+A')^2}{R_+}+\frac{(V'-A')^2}{R_-}\right] +&\nonumber\\
&\ \ \ +\frac{4 x^2Z(\lambda ) e^{-2a} q^2}{f}A^2 \ , & 
\label{eq:feq}
\end{align}
\end{subequations}
where 
\be \label{eq:Rpmdefapp}
 R_\pm = \sqrt{1-\frac{e^{-2 a} w(\lambda )^2 \left(A'\pm V'\right)^2}{q^2}} \ 
\ee
where the primes denote derivatives with respect to $a$. The equations for the vectorial gauge field (the sum of Eqs.~\eqref{eq:pluseom} and~\eqref{eq:minuseom}) and the equation for the blackening factor $f$ can be integrated once, giving
\begin{align}
\label{eq:Vint}
&\frac{e^{2a} V_f(\lambda ) w(\lambda )^2 }{2 q }\left[\left(\frac{1}{R_+}+\frac{1}{R_-}\right)V'+\left(\frac{1}{R_+}-\frac{1}{R_-}\right)A'\right] = c_V& \\
& \frac{e^{2a} x V_f(\lambda ) w(\lambda )^2 }{2 q }\left[\left(\frac{1}{R_+}+\frac{1}{R_-}\right)\left(VV'+AA'\right)+\left(\frac{1}{R_+}-\frac{1}{R_-}\right)\left(AV'+VA'\right)\right] = & \nonumber\\
&= \frac{f'e^{4a}}{q} + c_f, &
\label{eq:fint}
\end{align}
where $c_V$ and $c_f$ are the constants of integration. To obtain the latter equation here, one also needs to use the equations for the gauge fields.

The equations of motion are unchanged under the following two transformations:
\begin{align}
\label{eq:scaletransf}
 &a \mapsto a + \log \Lambda_a \ , \qquad V \mapsto \Lambda_a V\ , \qquad A \mapsto \Lambda_a A\ ; &\\
\label{eq:ftransf}
 &f \mapsto \Lambda_f f \ , \qquad q \mapsto \sqrt{\Lambda_f} q \ , \qquad V \mapsto \sqrt{\Lambda_f} V\ , \qquad A \mapsto \sqrt{\Lambda_f} A\ ; &
\end{align}
where $\Lambda_a$ and $\Lambda_f$ are positive constants. The former symmetry corresponds to the overall scale transformation discussed also in the main text. Indeed, the metric Eq.~\eqref{eq:metric} is unchanged under simultaneously applying Eq.~\eqref{eq:scaletransf} and the scaling $x_\mu \mapsto x_\mu/\Lambda_a$ of all spacetime coordinates. The latter symmetry is similarly identified as the separate scaling of the time coordinate. This symmetry is fixed by requiring that the blackening factor $f \to 1$ at the boundary so that the speed of light equals one (as it should since we are using natural units).

The background solutions are found numerically by shooting towards the boundary from the horizon $a=a_h$, where the blackening factor vanishes, $f(a_h)=0$. The gauge fields also vanish at the horizon. We initially fix the freedom related to the symmetries Eqs.~\eqref{eq:scaletransf} and~\eqref{eq:ftransf} by setting $a_h=0$ and $f'(0)=1$. The desired values for the energy scale and asymptotic value of $f$ are obtained thereafter by applying these mappings with well chosen $\Lambda_a$ and $\Lambda_f$. Following this, the regular solutions depend on three parameters; the natural choice for these parameters appears to be $\lambda_h = \lambda(0)$, $V_h' =V'(0)$, and $A_h'=A'(0)$. This however leads to regularity conditions at the horizon which cannot be solved analytically. One can try to bypass this issue by using instead the quantities
\be
 Q_V = \frac{V'(0)}{q(0)} \ , \qquad  Q_A = \frac{A'(0)}{q(0)} \ ,
\ee
which is actually equivalent to fixing $q$ at the horizon as opposed to $f'$. This choice is however not optimal because it turns out that the mapping from these variables to the chemical potentials is not monotonic in the region that we will be interested in. We find that this can be avoided by rather using the ``charges'' $n_{V/A}$ defined through
\be
 Q_V = \frac{n_V}{w(\l_h)\sqrt{V_f(\l_h)^2w(\l_h)^2+n_V^2}} \ , \qquad  Q_A 
 = \frac{n_A}{w(\l_h)\sqrt{V_f(\l_h)^2w(\l_h)^2+n_A^2}} \ .
\ee
The motivation for these definitions is that in the absence of an axial gauge field, $n_V$ is indeed (up to a normalization coefficient) the quark number density. 

Therefore the final input parameters for our numerical solutions are $\l_h$, $n_V$, and $n_A$. We introduce a small cutoff $\epsilon$ and start the numerical evolution of the equations of motion at $a=\epsilon$, with boundary conditions given by analytic horizon expansions of the regular solution obtained from the equations of motion\footnote{Starting directly at $a=0$ is not possible due to singular behavior of the system at this point.}. Note that we follow~\cite{Jarvinen:2011qe,Alho:2012mh}, choosing the sign of $q$ so that it is always negative. Moreover, the physically relevant branch has $\l'(a)<0$.

Finally after the numerical solution has been obtained, we use the symmetry Eq.~\eqref{eq:ftransf} to set $f=1$ at the boundary, $a \to \infty$. Observables can then be computed as follows. The  UV scale factor $\Lambda$, which appears in the near boundary expansions of $\l$ and the geometry~\cite{Jarvinen:2011qe}, can be computed by using
\be \label{eq:Lambdadef}
\Lambda = \lim_{a\to\infty}\frac{1}{\ell}\exp\left[a-\frac{8}{9 v_1\lambda(a) }+\left(\frac{23}{36}-\frac{16v_2}{9v_1^2}\right) \log \left(\frac{9 v_1 \lambda(a)}{8}\right)\right]
\ee
where the coefficients are defined through the weak coupling expansion of the effective potential
\be
 V_\mathrm{eff}(\l) = V_g(\l)-x V_f(\l) = \frac{12}{\ell^2}\left[1+v_1 \l +v_2 \l^2 +\mathcal{O}\left(\l^3\right)\right] \ .
\ee
The temperature and the vectorial chemical potential are then computed from
\be
 T = \frac{1}{4\pi}\left|\frac{f'(0)}{q(0)}\right| \ , \qquad \mu_V = \lim_{a\to \infty} V(a) \ .
\ee
The definition of the axial chemical potential depends on the choice for the function $Z(\l)$ and is discussed in App.~\ref{sec:ax_asym}. Finally, the dimensionless output is then given by the ratios $T/\Lambda$, $\mu_V/\Lambda$, and $\mu_A/\Lambda$.

\section{Axial gauge field asymptotics}\label{sec:ax_asym}
In this Appendix, we describe the asymptotic behaviour of the background axial gauge field, $A$, along with its associated fluctuation, $\delta A$, which are needed in order to extract the associated axial chemical potential, $\mu_{A}$ and axial magnetic field, $B_{A}$ respectively. We begin by assuming the Ansatz 
\begin{equation}
   \lim_{a\to\infty} A^i(a)\sim\mu^{i}_{A}e^{-\Delta_i a}a^{p_i}
   \bigg[1+\frac{1}{a}\Big(k^{i}_1+k^{i}_{11}\log{a}\Big)+\frac{1}{a^2}\Big(k^{i}_2+k^{i}_{22}\log{a}\Big)\bigg],
   \label{eq:ax_asym}
\end{equation}
where the $i$ superscript corresponds to the choice of the potential $Z(\lambda)$ and $\mu_{A}$ is the axial chemical potential. Then, by expanding either Eq.~\eqref{eq:pluseom} or Eq.~\eqref{eq:minuseom} in $1/a$ at the boundary, one can calculate $\Delta^i, p^i$ and the various $k^i$. With these coefficients in hand, we can read off the value of the axial chemical potential for each case 

\begin{align}
    \mu^{1}_{A}&=\mathcal{C}_1\lim_{a\to\infty}\frac{A^{1}(a)}{e^{-\Delta_1 a}a^{p_1}\Big[1+\frac{1}{a}\Big(k_1^1+k^1_{11}\log{a}\Big)\Big]}\,,
    \\\mu^{2}_{A}&=\mathcal{C}_2 \lim_{a\to\infty}\frac{A^{2}(a)}{a^{p_2}\Big[1+\frac{1}{a}\Big(k_1^2+k^2_{11}\log{a}\Big)+\frac{1}{a^2}\Big(k_2^2+k^2_{22}\log{a}\Big)\Big]}\,,
    \\\mu^{3}_{A}&=\lim_{a\to\infty}\frac{A^{3}(a)}{1+\frac{1}{a}\Big(k_1^3+k^3_{11}\log{a}\Big)+\frac{1}{a^2}\Big(k_2^3+k^3_{22}\log{a}\Big)}\,,
    \label{eq:muAc3}
\end{align}
where the various coefficients are listed in Table.~\ref{table:ax_coeffs}. Notice that for choice 3 in Eq.~\eqref{eq:muAc3} the chemical potential is directly related to the boundary value of the gauge field and the correction term in the denominator only improves convergence as $a \to \infty$. For the other choices, the correction term is required to obtain a finite value. Notice also that the normalization of the correction cannot be determined unambiguously for the choices 1 and 2, which is reflected in a similar normalization ambiguity of the chemical potentials. This is why we have included the coefficients $\mathcal{C}_i$ in the formulas. For the numerical results presented in the text, we used $\mathcal{C}_i = 10^{p_i}$. This was motivated by the fact that the value of $a$ at which the near boundary expansions start to converge is around $a=10$, so the coefficients cancel the large numerical factors arising from the power corrections $a^{p_i}$ at this point.

\begin{table}[h!]
	
	\begin{tabular}{ |p{0.1cm}|p{1.55cm}|p{1.25cm}|p{1.75cm}|p{1.4cm}|p{1.7cm}|p{1.8cm}|}
		\hline
		$i$ & $\Delta_i$ & $p_i$ & $k_1^i$ & $k_{11}^i$  & $k_2^i$ & $k_{22}^i$ \\
		\hline
		\hline
		1 & $-0.47048$ & $2.13839$ & $9.23171 \newline- 2.13839\mathcal{B}$ & $0.831595$ & neg. & neg. \\
		\hline
		2 & $0$ & $4.933$ & $25.5208 \newline- 4.993\mathcal{B}$ & $1.94172$ & $263.614 \newline- 103.847\mathcal{B} \newline+ 
 12.465\mathcal{B}^2$ & $38.4996 \newline- 9.69504\mathcal{B}$ \\
		\hline
		3 & $0$ & $0$ & $-2.59291$ & $0$ & $6.10334\newline - 2.59291\mathcal{B}$ & $1.008351$ \\
		\hline
	\end{tabular}
	\caption{Coefficients defining the asymptotic behaviour of the axial gauge field, $A(a)$. The quantity $\mathcal{B}\equiv\log\frac{l}{\Lambda_a}$ is background-dependent. For choice 1, we neglect the coefficients $k^{2}_{2},\,k^{2}_{22}$.}
	\label{table:ax_coeffs}
\end{table}

We note that $\Delta_1\equiv\Delta$ is indeed the anomalous dimension associated with the axial chemical and must therefore obey an additional constraint; as was pointed out in \cite{Jimenez-Alba:2014iia,Weitz}, for $\Delta<-1$, the axial counterpart of Eq.~\eqref{eq:conscurrent} would become an irrelevant operator, thereby destroying the desired $\text{AdS}_5$ asymptotics. Consequently, for choice 1 of potential, we must require $Z_0<0.906$. Referring to the discussion in App.~\ref{app:z_anal}, for choice 1 with $c_1=c_2=0$, one obtains a value of $Z_0>0.906$, thus necessitating the inclusion of a non-zero $c_1$.

Following an asymptotic analysis of Eqs.~\eqref{eq:Lfluct} and \eqref{eq:Rfluct}, we see that the fluctuation of the axial gauge field possesses the same behaviour in the UV as its background counterpart. We may thus define the axial magnetic fields 
\begin{align}
    B^{1\,j}_{A}-\mu^{1}_{A}\omega^j&=-\epsilon^{zjk}\mathcal{C}_1\lim_{a\to\infty}\frac{\delta \hat{A}_k^{1}(a)}{e^{-\Delta_1 a}a^{p_1}\Big[1+\frac{1}{a}\Big(k_1^1+k^1_{11}\log{a}\Big)\Big]}\,,
    \\B^{2\,j}_{A}-\mu^{2}_{A}\omega^j&=-\epsilon^{zjk}\mathcal{C}_2 \lim_{a\to\infty}\frac{\delta \hat{A}_k^{2}(a)}{a^{p_2}\Big[1+\frac{1}{a}\Big(k_1^2+k^2_{11}\log{a}\Big)+\frac{1}{a^2}\Big(k_2^2+k^2_{22}\log{a}\Big)\Big]}\,,
    \\B^{3\,j}_{A}-\mu^{3}_{A}\omega^{j}&=-\epsilon^{zjk}\lim_{a\to\infty}\frac{\delta \hat{A}^{3}_k(a)}{1+\frac{1}{a}\Big(k_1^3+k^3_{11}\log{a}\Big)+\frac{1}{a^2}\Big(k_2^3+k^3_{22}\log{a}\Big)}\,,
\end{align}
where the $j$ index refers to the spatial index of the axial magnetic field.

{\small

\bibliographystyle{utphys}
\bibliography{anomvqcd.bib}
}

\end{document}